\documentclass{article}
\usepackage[T1]{fontenc} 
\usepackage[utf8]{inputenc} 
\usepackage{ismir,amsmath,cite,url}
\usepackage{graphicx}
\usepackage{color}
\usepackage{array} 
\usepackage{booktabs}
\usepackage{lineno}
\usepackage{multirow}
\usepackage{multicol}
\usepackage{makecell}
\usepackage{array}    
\usepackage[table]{xcolor} 
\definecolor{lightgray}{gray}{0.9}
\usepackage{siunitx} 
\usepackage{enumitem}
\usepackage{float}
\title{Automatic Detection of Moral Values in Music Lyrics}

\multauthor
{Vjosa Preniqi$^1$ \hspace{1cm} Iacopo Ghinassi$^1$ \hspace{1cm} Julia Ive$^1$} { \bfseries{Kyriaki Kalimeri$^2$ \hspace{1cm} Charalampos Saitis$^1$}\\
 $^1$ Centre for Digital Music, Queen Mary University of London, London, UK\\
$^2$ ISI Foundation, Turin, Italy\\
{\tt\small \{v.preniqi, i.ghinassi, j.ive, c.saitis\}@qmul.ac.uk, kyriaki.kalimeri@isi.it}
}

\def\authorname{V. Preniqi, I. Ghinassi, J. Ive, K. Kalimeri, and C. Saitis}

\begin{document}

\maketitle
\begin{abstract}
Moral values play a fundamental role in how we evaluate information, make decisions, and form judgements around important social issues. The possibility to extract morality rapidly from lyrics enables a deeper understanding of our music-listening behaviours. 
Building on the Moral Foundations Theory (MFT), we tasked a set of transformer-based language models (BERT) fine-tuned on 2,721 synthetic lyrics generated by a large language model (GPT-4) to detect moral values in 200 real music lyrics annotated by two experts.
We evaluate their predictive capabilities against a series of baselines including out-of-domain (BERT fine-tuned on MFT-annotated social media texts) and zero-shot (GPT-4) classification. 
The proposed models yielded the best accuracy across experiments, with an average F1 weighted score of 0.8. This performance is, on average, 5\% higher than out-of-domain and zero-shot models. When examining precision in binary classification, the proposed models perform on average 12\% higher than the baselines.
Our approach contributes to annotation-free and effective lyrics morality learning, and provides useful insights into the knowledge distillation of LLMs regarding moral expression in music, and the potential impact of these technologies on the creative industries and musical culture.
\end{abstract}
\section{Introduction}\label{sec:introduction}
\begin{figure}
 \centerline{
 \includegraphics[width=1.\columnwidth]{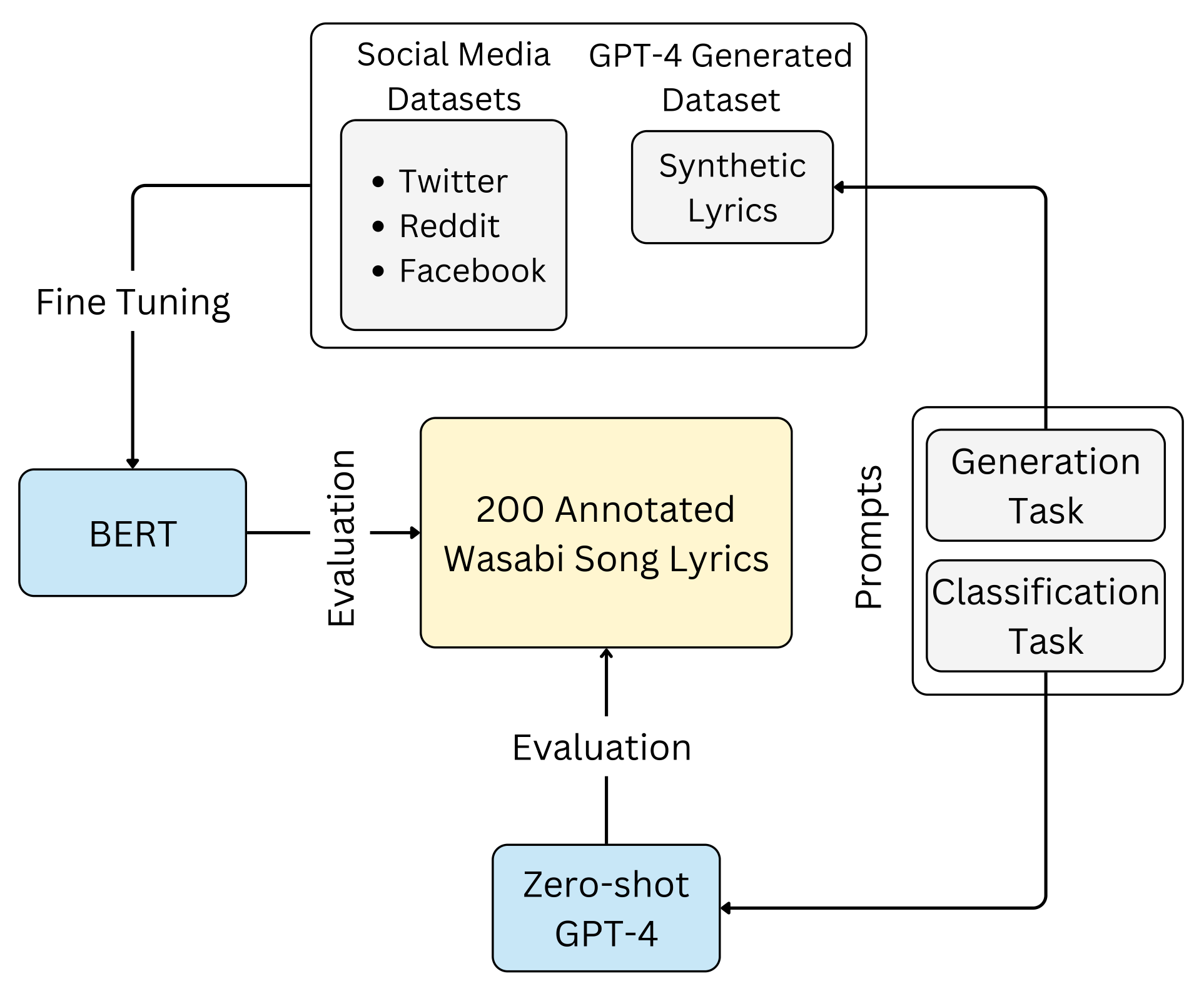}}
 \caption{Model Structure for predicting Moral Foundations (MFT) in Lyrics, fine-tuned on out-of-domain social media data, and synthetically generated lyrics with GPT-4. }
 \label{fig:Overall_popeline}
\end{figure}

Lyrics play a crucial role in how we experience music, affecting our emotions and actions. Positive lyrics can motivate and elevate listeners, whereas negative or aggressive content in songs may negatively impact mood and behaviour \cite{ballard1995immediate}. 
Social, political, and cultural issues, such as racial inequality and gender discrimination, are often reflected in the music lyrics of their time \cite{frith1981sound,betti2022large}.
Songs that feature in successful campaigns typically include uplifting melodies and lyrics that reflect the ideals of a nation, representing values of optimism and progress towards a better future \cite{dewberry2014music}.
Moral rhetoric in lyrics has been used to advocate for what is perceived to be a necessary societal change \cite{kizer1983protest}, promote peace and unity \cite{adebayo2017vote}, and raise awareness for marginalised groups \cite{sellnow1999music}. These narratives are closely related to moral judgements and beliefs, yet their relationship to music listening behaviors has received limited attention by music scientists. 

In the field of Music Information Retrieval (MIR), lyrical content analysis has  focused primarily on genre classification \cite{mayer2011musical}, mood prediction \cite{delbouysmusic}, emotion dynamics \cite{song2023modeling}, and lyrics-to-audio alignment \cite{vaglio2020multilingual, masclef2021user}. 
Recent works have elaborated on less attended psychological characteristics of music lyrics, including moral valence.
For example, insights into personal values and personality traits derived from lyrics can enhance various MIR tasks, including genre classification, audio tagging, and music recommendations \cite{kim2020butter}.
Preniqi and colleagues \cite{preniqi2023soundscapes} showed that moral valence extracted from lyrics can to some extent predict listeners’ moral values, in some cases more accurately than audio features. 
The possibility to extract morality rapidly from lyrics can enable a deeper understanding of our music listening behaviours.

Inferring moral values from song lyrics is a complex natural language processing (NLP) task from the start due to the subjectivity of our perceptions and interpretations. The progress is further hindered by the lack of annotated lyrics for training new or fine-tuning pre-trained models, and for benchmarking. Using models fine-tuned with out-of-domain annotated texts (e.g., from social media \cite{preniqi2024moralbert,guo2023data}) to predict moral values in music lyrics faces significant challenges due to the unique structure of lyrics compared to other textual forms (e.g., greater use of repetition, metaphor, imagery, and other poetic devices).

In light of the above, we investigate the novel task of automatic detection of moral values in music lyrics using an integrated approach that leverages the strengths of two distinct NLP technologies. Specifically, we leverage the generative capabilities of GPT-4 (Generative Pre-trained Transformer) to create morally nuanced synthetic lyrics---a process required only once---and employ BERT (Bidirectional Encoder Representations from Transformers), which demands fewer computational resources, to learn from the synthetic data structure.

Following recent related work \cite{preniqi2022more,preniqi2023soundscapes,preniqi2024moralbert}, we operationalize morality drawing on Haidt and Graham's Moral Foundations Theory \cite{haidt2007morality}, which outlines five core moral traits, or foundations, divided into ``virtue'' and ``vice'' based on moral polarity: \textit{Care} and \textit{Harm}, \textit{Fairness} and \textit{Cheating}, \textit{Loyalty} and \textit{Betrayal}, \textit{Authority} and \textit{Subversion}, \textit{Purity} and \textit{Degradation}. 
We developed a corresponding set of 10 single-label classification models, each customized to predict the presence or absence of one moral value in lyrical text.
MFT is a straightforward yet comprehensive model for understanding moral values, uniquely characterized by well-developed term dictionaries \cite{hoover2018moral}.

We present a dataset of 200 real song lyrics human-annotated with MFT. To the best of our knowledge, this is the first such dataset. It serves as the basis for evaluating our proposed method. We make the real and synthetic lyrics datasets, and the paper code fully available via a GitHub repository.\footnote{https://github.com/vjosapreniqi/ismir-mft-values}

We report a comprehensive comparison of the proposed models against BERT fine-tuned with out-of-domain human-annotated moral text data and zero-shot classification with GPT-4. Figure \ref{fig:Overall_popeline} summaries the overall pipeline of this work.
The proposed models yielded the best accuracy across experiments, with an average F1 weighted score of 0.8. This performance is, on average, 5\% higher than out-of-domain and zero-shot models. When examining precision in binary classification, the proposed models perform on average 12\% higher than the baselines. Our approach contributes to annotation-free lyrics morality learning, and provides useful insights into the knowledge distillation of large language models such as GPT-4 regarding moral expression in music.

\section{Related Work}
The field of music and moral expression has received limited attention. However, recent studies have shown a link between an individual's moral values and their preferences for lyrics and music, suggesting significant implications for tailoring personalisation in streaming services \cite{preniqi2021modelling, preniqi2023soundscapes, preniqi2022more}. Further research has delved into how moral values and lyrical preferences manifest within specific music communities. For example, Messick and Aranda \cite{messick2020role} demonstrated that moral values could explain a unique and significant portion of the variance in lyrical preferences among fans of different metal music sub-genres. 

Given the understanding that verbal expressions more effectively convey morality than non-verbal forms \cite{preniqi2022more, kalimeri2019predicting}, initial studies introduced lexicons \cite{Araque2020, hopp2021extended} as an extension of  Moral Foundations Dictionary (MFD)~\cite{Graham2009}
for identifying words and lemmas that accurately depict moral foundations.
More recent studies focused on examining moral values in texts using human-annotated social media datasets\cite{hoover2020moral, trager2022moral, beiro2023moral}, and introducing more advanced Natural Language Processing (NLP) approaches to detect moral dimensions in textual content\cite{guo2023data, preniqi2024moralbert}. Trager et al.\cite{trager2022moral} introduced baseline models for predicting moral values, employing a pre-trained BERT model fine-tuned on the Moral Foundation Reddit Corpus.  Guo et al. \cite{guo2023data} proposed a multi-label model for predicting moral values with Twitter and news data, incorporating the domain adversarial training framework suggested by Ganin et al.~\cite{ganin2015unsupervised} to align multiple datasets and generalise for out-of-domain predictions.
A similar approach was taken by Preniqi et al. \cite{preniqi2024moralbert} in predicting moral values in different social media domains.

However, a main challenge that persists is the ability of these models to generalise across various domains. Lisco and colleagues\cite{liscio2023does} demonstrated that text classifiers perform better when domains are similar.
This poses a major obstacle when predicting morality in lyrics because there is no prior study that has presented an annotated lyrics dataset with moral values. Further, manually annotating extensive text demands substantial time, resources, and deep understanding of Moral Foundations Theory (MFT).

To overcome these limitations, we employ GPT-4, an advanced LLM, to generate lyrics infused with various moral undertones, which helps in fine-tuning a moral classifier. This minimises the need for laborious manual annotation of extensive lyric databases, enabling us to utilise a smaller, human-annotated dataset to validate the effectiveness of knowledge distilled from GPT-4.
The capacities of LLMs for music tasks are being actively explored for the moment. Doh et al.\cite{doh2023lp} similarly employed a large language model such as GPT-3 for generating pseudo captions from tags to mitigate the problem of data scarcity in the field of automatic music captioning. 
While Zhang et al \cite{zhang2024syllable} evaluated the quality and correctness of generated music lyrics via GPT-3. Sawicki et al. \cite{sawicki2023power} investigated the possibility of using GPT-3 models to generate high-quality poems in a specific author’s style while suggesting that GPT-3 can be a useful tool in assisting authors. 

\section{Method}
\subsection{Human-Annotated Lyrics}
\begin{figure}
 \centerline{
 \includegraphics[width=1.\columnwidth]{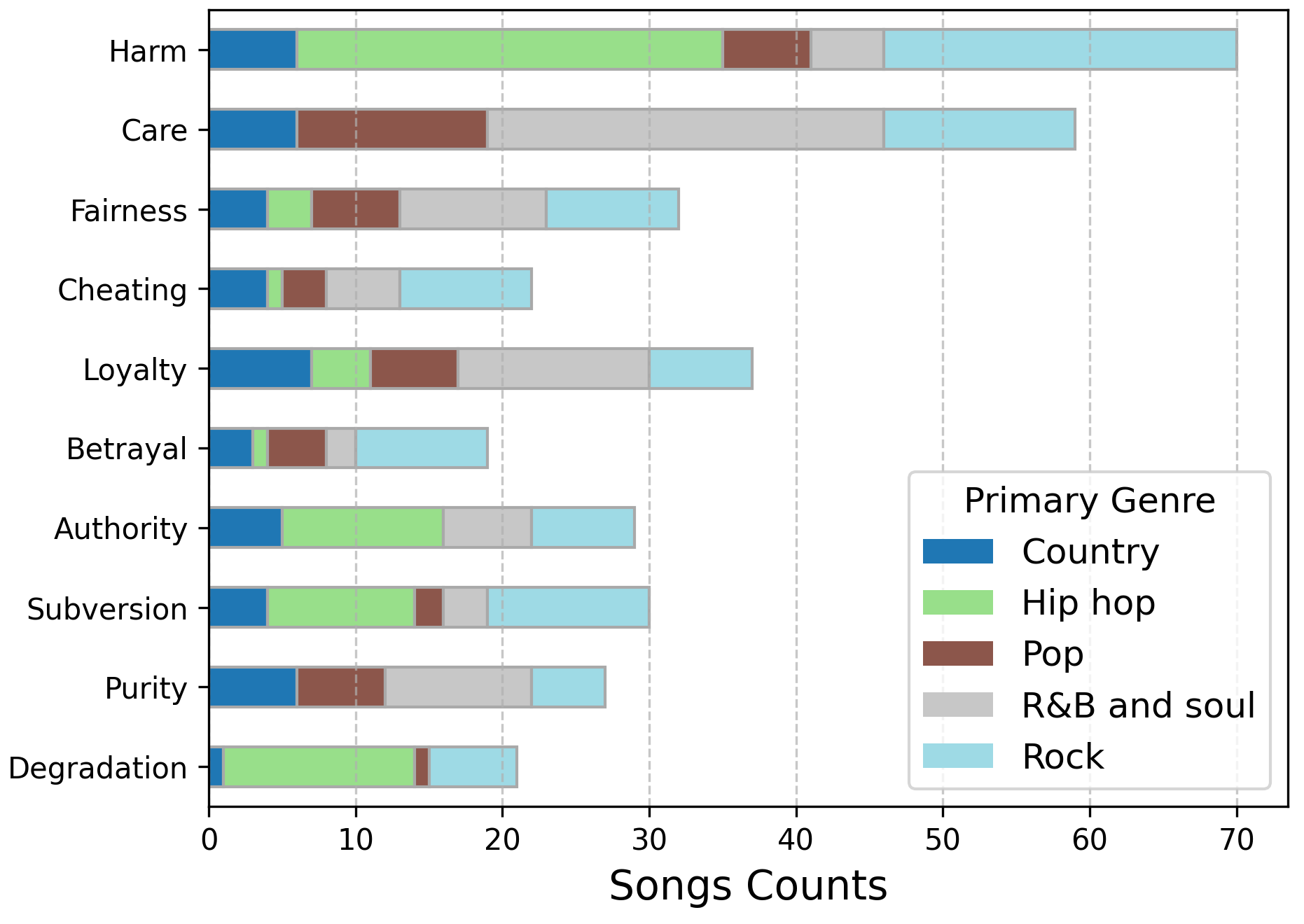}}
 \caption{Distribution of Moral Foundations in 200 song lyrics dataset annotated by human annotators with genre proportions for each moral value.}
 \label{fig:mft_values_with_genre_chart}
\end{figure}

For this work, we annotated 200 song lyrics, categorising them into 10 different moral foundations. This annotation process was conducted by two skilled annotators: the lead author of this study and an external researcher with a background in music and sound design, both of whom agreed to contribute. Before starting, the annotators were informed about their participation rights, including the option to discontinue their involvement at any point. 
Each annotator was assigned with 125 songs for annotation. To evaluate the agreement between annotators, 50 songs were annotated by both annotators. The inner-annotator agreement was assessed using Cohen’s kappa coefficient for each moral label. This resulted in an almost perfect agreement \cite{landis1977measurement} with an average score of 0.86 across all moral categories identified within the lyrics of the chosen songs. 
We selected the songs for the moral values annotation from the Wasabi Dataset \cite{meseguer2017wasabi}, known for its extensive collection of 2 million songs including lyrics, artist gender, and musical genre among other data. This dataset spans over five decades, enabling the selection of songs from various eras. The process of selecting the songs involved a semi-random approach, with efforts made to retain the distribution of genres, and the timeline of song releases as found in the original dataset.
Among the 200 songs annotated for moral values, 18 were from the 60s and 70s, 78 from the 80s and 90s, and 116 from the post-2000 era. The chosen songs represented a balanced mix of genres including Rock, Pop, Hip-Hop, R\&B, Soul, and Country. Figure \ref{fig:mft_values_with_genre_chart} depict the distribution of Moral values in the human-annotated song lyrics with the proportion of genre for each MFT value.

\subsection{Predicting Morality in Lyrics with Domain Adaptation}
Initially, we tried to predict moral values in lyrics by fine-tuning a BERT model with out-of-domain social media data, following the approach used by Preniqi et al. \cite{preniqi2024moralbert}.
We utilised 20,628 tweets from the Moral Foundation Twitter Corpus (MFTC)  \cite{hoover2020moral}; 13,995 posts from the Moral Foundations Reddit Corpus (MFRC) \cite{trager2022moral}; 1,510 posts from Facebook vaccination dataset \cite{beiro2023moral}. 
Preniqi's and other work have demonstrated that predicting moral values using a single-label approach---predicting one MFT value at a time---results in higher accuracy \cite{trager2022moral,preniqi2024moralbert}. Informed by these findings, we developed a set of single-label classification models tailored to predict individual moral foundations in lyrics.

As a baseline model, we apply a similar approach to the MoralBERT \cite{preniqi2024moralbert}. We identify the polarities (virtues and vices) of moral foundations, as opposed to just identifying the mere presence or absence of moral values. We incorporate the domain adversarial method aiming to improve the models' ability to generalise effectively in predicting moral values in lyrics \cite{guo2023data, preniqi2024moralbert}.
Adopting this model, we start by deriving a domain invariant representation $h$ from the BERT CLS embedding $e$:
\begin{equation*}
    h=W_{inv}e
\end{equation*}
where $W_{inv} \in \mathcal{R}^{768 \times 768}$ is a learnable matrix. Next, we calculate moral values predictions $\hat{y_m}$ using:

\begin{equation*}
\label{eqn:prediction}
    \hat{y_m}=Softmax(W_1(ReLU(W_2h)))
\end{equation*}
with $W_1 \in \mathcal{R}^{768 \times 768}$, $W_2 \in \mathcal{R}^{768 \times c}$ representing 2 learnable matrices, $c$ being the number of classes, $ReLU$ is the rectified linear unit activation function and $Softmax$ is the normalised exponential function.
A domain classification head is also included for obtaining domain predictions $\hat{y_d}$: 
\begin{equation*}
\hat{y_d} = Softmax(W_3(ReLU(W_4h)))
\end{equation*}
with $W_3 \in \mathcal{R}^{768 \times 768}$, and $W_4 \in \mathcal{R}^{768 \times d}$ learnable matrices and with d being the number of domains in the training set.
The main rationale of the adversarial network is increasing the loss from the domain head while minimising the loss from the moral values prediction. Hence, the model is ``forced'' to learn domain-invariant representations. This is achieved by integrating a gradient reversal layer before the domain classification head, while using standard training for minimising moral prediction loss. Cross-entropy (\(CE\)) loss is used for both the moral and domain classification heads. The final loss is expressed as:
\begin{equation*}
L = CE(\hat{y_m}, Y_m) - CE(\hat{y_d}, Y_d) + L_{norm} + L_{rec}
\end{equation*}
with and \(Y_m\) and \(Y_d\) as the ground truth for moral values and domain, respectively.
Two regularisation terms from \cite{guo2023data} are added: L2 norm regularisation and reconstruction loss:
\begin{align*}
    L_{norm}=||W_{inv}h-I||^2, \quad
    L_{rec}=||W_{rec}h-e||^2
\end{align*}
similar to $W_{inv}$ (defined above), $W_{rec} \in \mathcal{R}^{768 \times 768}$ is also a learnable matrix and \(I\) is the identity matrix. These regularization losses are combined with moral and domain classification losses. The regularization terms are not applied when training MoralBERT on a single domain (e.g., when trained on just synthetic lyrics).

The binary setting we use implies the model should learn from highly unbalanced datasets, where the neutral label (negative class) is far more represented than the single moral value to be predicted in each instance (positive class). To address the class imbalance, we employed two methods. First, weights are assigned to classes\cite{king2001logistic}:
\begin{equation*}
    weight_c=\frac{N-N_c}{N}
\end{equation*}
where $N$ is the total training samples and $N_c$ is the count of samples per class $c$.
Second, similar to \cite{koshorek2018text}, we employed a separate threshold $\theta_v$ for each moral value $v$, so that we use $\hat{y_m}$ to obtain the final prediction $\hat{m}$:
\begin{equation*}
\hat{m}=\begin{cases}
1 &\text{if $\hat{y_m}>\theta_v$}\\
0 &\text{otherwise}
\end{cases}
\end{equation*}
with $\hat{m}=1$ indicating the moral value is present in the lyrics and $\hat{m}=0$ indicating it is not.
The optimal value $\theta_v$ for each moral value $v$ was found by optimizing for binary F1 during training, searching in the search space 0.05 to 0.95 with a step of 0.05. The models were trained for 20 epochs using a single Nvidia T4 GPU, a learning rate of 5e-5, and the Adam optimiser for all MoralBERT experiments.

\subsection{Synthetic Lyrics Generation for Moral Assessment}



There is a growing interest in knowledge distillation from large pre-trained language models via synthetic text generation~\cite{ye-etal-2022-zerogen}. Here we apply a similar knowledge distillation approach by utilising GPT-4 for synthetic lyrics generation. This method eliminates the need to collect real-life data, which is often difficult to gather for a specific NLP task and with a specific input distribution \cite{he2022generate}.
Initially, we assessed GPT-4's familiarity with Moral Foundation Theory \cite{Graham2009}, confirming its fundamental understanding of moral values. 
We tasked GPT-4 with generating lyrics by formulating a prompt, as follows: \\
\textbf{Prompt}: \textit{You are an assistant to a songwriter, you need to assist in writing lyrics related to the Moral foundations described in the Moral Foundation
Theory. Given the} \{\texttt{Moral Foundations Tags}\} \textit{, which represent} \{\texttt{Description Tags}\}\textit{, write original lyrics of a song expressing these moral foundations. DO NOT directly mention these moral foundations. DO NOT explicitly talk about morality. Write it in the style of} \{\texttt{Artist Tags}\}.

We assigned a ``role'' (songwriter assistant) for the model and provided three types of ``input tags''. The  \{\texttt{Moral Foundations Tags}\}  comprise any of the 10 moral values. The resulting lyrics can represent 1, 2, or 3 moral values. We determined this based on the moral combinations observed in our human-annotated lyrics dataset. 
The \{\texttt{Description Tags}\} represent fundamental concepts of each moral value.
The \{\texttt{Artist Tags}\} represent the names of artists whose styles we employ to diversify the lyrics. Initially, we intended to commence the lyrics generation task solely using moral categories and genres as tags. However, we observed that the lyrics were more uniform and generic compared to when we incorporated the artist's style. To tailor the lyrical style using various artists, we employed MusicOSet \cite{silva2019musicoset}, a collection of musical elements (e.g., music, albums, artists, genres and popularity) suitable for music data mining. To capture the nuances of different genres, we organized the artists according to their popularity and grouped them into prevalent genres like Rock, Pop, Country, Hip Hop, R\&B, Soul, Folk, Blues, and Jazz. These genres align very closely with those in the song lyrics we selected for human annotations. We chose to utilise this dataset because it offers detailed data on artist genres and sub-genres, as well as an artist popularity metric that we employ in developing lyric styles. We acquired a dataset comprising 2,721 artificially generated lyrics, each aligned with moral categories similar to our human-annotated lyrics dataset. On average, the generated lyrics had 146 words, with a total of 10,305 unique words across the synthetic lyrics dataset. 


\subsection{GPT-4 in Moral Classification Task}

In addition, we wanted to assess the capability of the 0-shot GPT-4 model in classifying morality in actual song lyrics while comparing it to our proposed model. To do so, we prompted the task as follows:

\textbf{Prompt}: \textit{You will be provided with song lyrics. The song lyrics will be delimited with \#\#\#\# characters. Classify each lyric into 10 Possible Moral Foundations as defined in Moral Foundation Theory The available Moral Foundations are:} \{\texttt{Moral Foundations Tags}\}.
\textit{The explanation of the moral foundations is as follows:}  
\{\texttt{Description Tags}\}.
\textit{This is a multi-label classification problem: where it's possible to assign one or multiple categories simultaneously. 
Report the results in JSON format such that the keys of the correct moral values are reported in a list}.

The song lyrics utilised for the GPT-4 model classification are the same as the ones annotated by human annotators. In this way, we can compare the human annotations with those of the model while assessing the general performance of GPT-4 for the classification task. 


\section{Experiments}

\begin{table*}[!ht]
\small
\centering
\begin{tabular}{@{}lcccc|cccc}
\toprule\toprule
\multicolumn{1}{l}{} & \multicolumn{4}{c|}{\textbf{F1 Scores Weighted Average}} & \multicolumn{4}{c}{\textbf{F1 Scores Binary}} \\
\midrule
\multicolumn{1}{l}{} & \multicolumn{1}{c}{\begin{tabular}[c]{@{}c@{}}MoralBERT\end{tabular}} & \multicolumn{1}{c}{ GPT-4 } & BERT SL & \begin{tabular}[c]{@{}c@{}}MoralBERT SL\end{tabular} & \multicolumn{1}{c}{\begin{tabular}[c]{@{}c@{}}MoralBERT \end{tabular}} & \multicolumn{1}{c}{ GPT-4 } & BERT SL & \multicolumn{1}{c}{\begin{tabular}[c]{@{}c@{}}MoralBERT SL\end{tabular}} \\
\midrule
Care & .80 ± .03 & .68 ± .03 & .81 ± .03 & \textbf{.83 ± .03} & .68 ± .05 & .64 ± .04 & .68 ± .05 & \textbf{.75 ± .04} \\
Harm & .68 ± .03 & \textbf{.75 ± .03} & .71 ± .03 & .70 ± .03 & .62 ± .05 & \textbf{.71 ± .04} & .63 ± .05 & .69 ± .04 \\
Fairness & .55 ± .03 & .73 ± .03 & .73 ± .03 & \textbf{.74 ± .03} & .30 ± .05 & .39 ± .06 & \textbf{.41 ± .06} & .38 ± .06 \\
Cheating & .84 ± .03 & .80 ± .03 & \textbf{.86 ± .02} & .69 ± .03 & .27 ± .09 & .16 ± .07 & \textbf{.52 ± .08} & .32 ± .06 \\
Loyalty & .69 ± .03 & .67 ± .03 & .77 ± .04 & \textbf{.79 ± .04} & \textbf{.38 ± .06} & .34 ± .06 & .21 ± .08 & .27 ± .09 \\
Betrayal & .81 ± .02 & .72 ± .03 & \textbf{.89 ± .02} & .84 ± .02 & .34 ± .07 & .31 ± .06 & \textbf{.40 ± .11} & .37 ± .08 \\
Authority & .77 ± .03 & .75 ± .03 & .77 ± .03 & \textbf{.84 ± .03} & \textbf{.45 ± .06} & .42 ± .06 & .35 ± .07 & .39 ± .09 \\
Subversion & \textbf{.80 ± .03} & .72 ± .03 & \textbf{.80 ± .03} & .71 ± .03 & \textbf{.44 ± .07} & .39 ± .06 & .40 ± .07 & .43 ± .06 \\
Purity & .77 ± .03 & .86 ± .02 & .89 ± .02 & \textbf{.90 ± .02} & .41 ± .06 & .56 ± .07 & .55 ± .08 & \textbf{.63 ± .08} \\
Degradation & .74 ± .03 & .81 ± .03 & .81 ± .03 & \textbf{.86 ± .03} & .34 ± .06 & \textbf{.40 ± .07} & .30 ± .07 & .32 ± .10 \\
\midrule
Average & .75 ± .03 & .75 ± .03 & \textbf{.80 ± .03} & \textbf{.80 ± .03} & .42 ± .06 & .43 ± .06 & .45 ± .07 & \textbf{.46 ± .07} \\
\bottomrule\bottomrule
\end{tabular}
\caption{F1 scores of prediction models with standard deviation estimated via 1,000 bootstraps. Weighted average scores account for both moral and non-moral (neutral) classes, while binary scores only for moral classes. SL = Synthetic Lyrics.}
\label{tab:MFT_Prediction_Models}
\end{table*}

\begin{figure*}
 \centerline{
 \includegraphics[width=1.9\columnwidth]{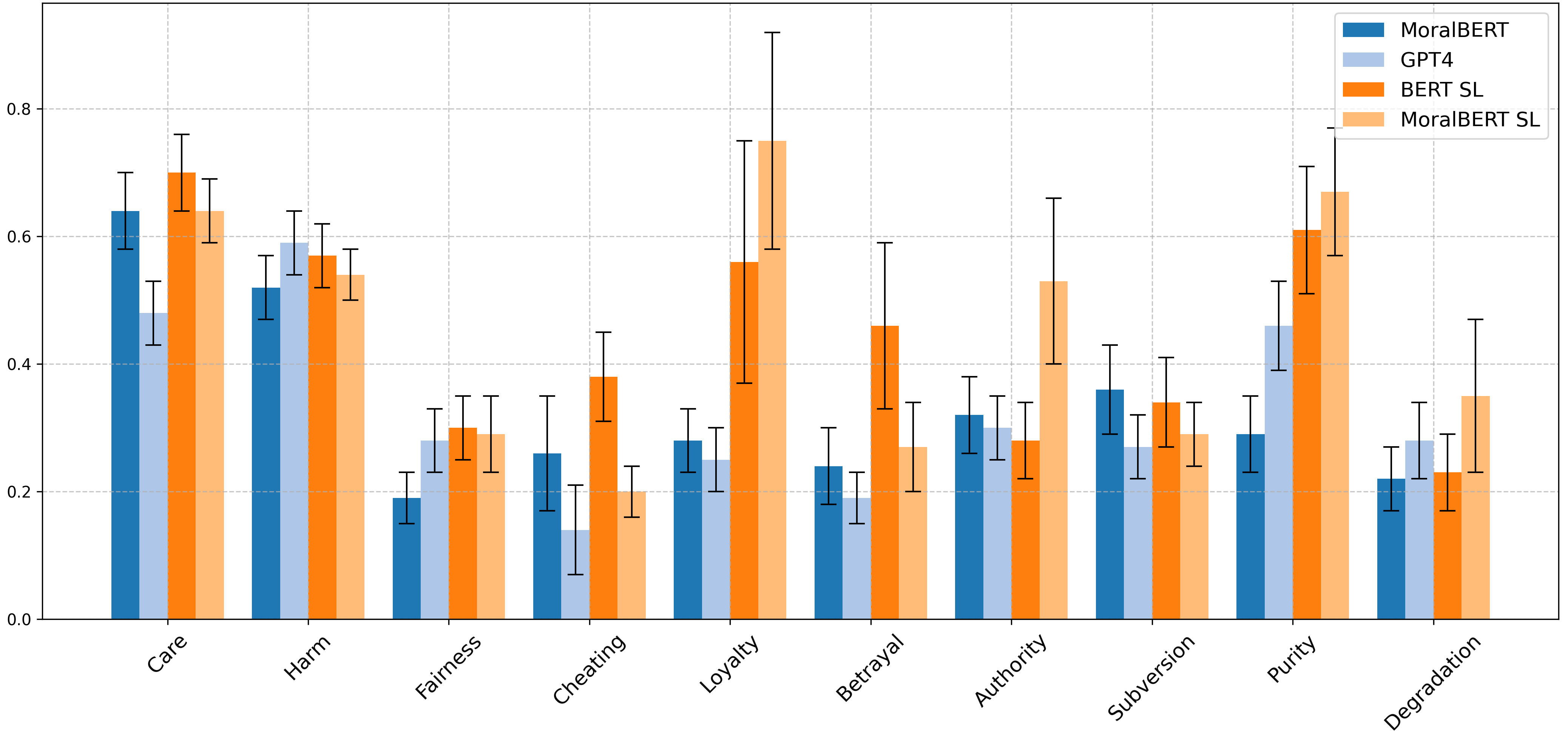}}
 \caption{Precision scores for binary classification with standard deviation estimated via 1,000 bootstraps.}
 \label{fig:Precision_binary_scores}
\end{figure*}

We started by analysing the MoralBERT technique \cite{preniqi2021modelling} and fine-tuned models using social media data from Twitter, Reddit, and Facebook. The total number of text records was 35,887. We found that 51\% of the texts were neutral and 49\% of them were labeled with one or more moral values. This indicated a significant skew towards neutral texts, which we addressed by adding the class weighting technique. After that, we evaluated the BERT models fine-tuned with only GPT-4 generated lyrics. We call these models "BERT SL". We also fine-tuned the models with a combination of out-of-domain social media data and the generated lyrics data which we call "MoralBERT SL". We used the Domain Adversarial module only when fine-tuning BERT with multiple domain data, including synthetic lyrics. When fine-tuning solely with synthetic lyrics, this module was not utilized. Lastly, we evaluated GPT-4's zero-shot classification capabilities against our models on the manually annotated song lyrics.

The results show that the models achieving the highest performance were BERT SL and MoralBERT SL. These models performed on average 5\% better across all moral values in terms of  F1 weighted score which accounts for both moral and non-moral prediction classes. While for the binary F1, these models were marginally better than GPT-4. For harm foundation, GPT-4 performed slightly better, possibly due to the synthetic lyrics' lack of natural variability when expressing this foundation. 
The fact that MoralBERT SL and BERT SL performances are similar to the one from GPT-4 for binary F1 is expected as the same latent knowledge of GPT-4 has been distilled into BERT by using the generated lyrics.
The improvements from MoralBERT SL and BERT SL are significant for what concerns weighted F1, suggesting that given the supervised setting of these models, they were also able to learn the higher prior probability of non-moral (e.g., neutral) instances, which generally outweigh moral instances. 
The same is evident if we look at Figure \ref{fig:Precision_binary_scores}, which compares the binary Precision scores of the various models. From the figure, it is evident that MoralBERT SL and BERT SL exhibit significantly higher Precision surpassing GPT-4 and MoralBERT by 12\% on average. These models, then, are often correct when labelling lyrics with moral values (even though results vary according to which moral value), while being more cautious in assigning a moral value, given the preponderance of neutral cases. For the evaluation metrics, we report the standard deviation estimated via Bootstrapping which is a statistical resampling technique used to estimate the variability of the metrics. We used 1,000 bootstraps which is typically sufficient to achieve a reasonable approximation of the standard deviation.
\begin{table*}[!ht]
\centering
\small
\renewcommand{\arraystretch}{1.5} 
\begin{tabular}{>{\centering\arraybackslash}m{2.3cm}>{\centering\arraybackslash}m{2.3cm}|>{\centering\arraybackslash}m{2.3cm}>{\centering\arraybackslash}m{2.3cm}>{\centering\arraybackslash}m{2.3cm}>{\centering\arraybackslash}m{2.3cm}}
\toprule\toprule
\textbf{Song Name} & \textbf{Artist} & \textbf{Human Annotations} & \textbf{MoralBERT} & \textbf{GPT-4} & \textbf{BERT SL MoralBERT SL} \\ \midrule
``Take This Heart of Mine'' & Foghat & Care, Purity & Care, Purity & Care, Loyalty & Care, Fairness, Purity \\ 
``Who's Cheatin' Who'' & Charly McClain & Cheating, Betrayal & Cheating, Betrayal, Loyalty, Purity & Cheating, Betrayal & Cheating, Betrayal \\ 
``Samurai Showdown'' & RZA & Harm, Authority & Harm, Betrayal, Authority, Purity & Harm, Loyalty, Authority & Harm, Authority \\ 
``Man In The Mirror'' & Mark Chesnutt & Care, Fairness & Fairness, Loyalty, Authority & Care, Fairness, Loyalty, Authority & Care, Fairness \\ \bottomrule\bottomrule
\end{tabular}
\caption{Examples of moral values detected in song lyrics by human annotators and model predictions.}
\label{tab:qualitative_evaluation}
\end{table*}

Our findings show that BERT-based models are still comprehensible with larger models such as GPT-4, when fine-tuned properly they can excel in specified tasks. GPT-4 demonstrated a very good performance even without any fine-tuning (zero-shot approach) which was anticipated given its state-of-art performance in multiple tasks and its training on an extensive amount of data. These models have been trained on diverse text sources such as Wikipedia, GitHub, chat logs, books, and articles \cite{brown2020language}, enabling them to comprehend language across various domains \cite{doh2023lp}. The earlier model, GPT-3, contains 175 billion parameters, a figure vastly greater than BERT’s 340 million parameters \cite{bosley2023we}. Such models demand significantly more computational resources than BERT models. In contrast, the BERT model is cost-free, easier to modify, and offers greater control over the models due to its open-source nature.  On the other hand, BERT models need fine-tuning, which presents its own challenges due to the necessity for manual labelling and data annotation. Therefore, a hybrid approach like the one we suggest offers an optimised solution that combines the best of both worlds.


Table \ref{tab:qualitative_evaluation} presents four song examples annotated for moral values by both human annotators and prediction models. These examples show that MoralBERT SL and BERT SL (not shown in the table as it shares the same outcomes as MoralBERT SL for these instances) aligned most closely with human moral assessments. From a general observation of the song lyrics that were annotated by humans and tested with these models, it was noted that MoralBERT and GPT-4  tend to assign more moral attributes per song while increasing their chances of correctly guessing moral labels but also misclassifying neutral ones. In contrast, models trained with synthetic lyrics more accurately identified neutral (non-moral) lyrics, aligning with the quantitative observations of the F1 weighted score.
Typically, human annotators did not assign more than three moral values per song. 
To control the number of assigned moral values per song, we adjusted the thresholds \cite{koshorek2018text} for our prediction models, ensuring optimal accuracy. When lacking ground truth data, a post-processing can be applied for cutting moral labels with lower probabilities.
Here we present only F1 and Precision scores. For further details, refer to the project's results page on GitHub.\footnote{https://github.com/vjosapreniqi/ismir-mft-values/tree/main/Results}

\section{Conclusion}
In this paper, we presented an integrated approach for the automatic detection of moral values in  lyrics. 
We created a synthetic lyrics dataset using GPT-4 which we used to fine-tune the BERT-base model alone (BERT SL) and in combination with out-of-domain social media corpora (MoralBERT SL). We introduced a  dataset of 200 song lyrics sourced from the WASABI dataset annotated for moral values by two experts, serving as the basis for evaluating our moral prediction models. We also assessed the performance of models trained with synthetic lyrics in comparison to those trained solely on social media data (MoralBERT) and a zero-shot GPT-4 classifier. We found that models trained with synthetic lyrics generally achieved significantly better binary Precision and higher weighted F1 scores compared to the GPT-4 classifier and MoralBERT, along with marginally better binary F1. 

Our research has some limitations. To begin with, the synthetic lyrics is created via GPT-4, a powerful model but not an open-source, which limits our control of the model.  We prompted GPT-4 to create unique lyrics in the style of various artists across different genres. Yet, adding musical composition details, lyrical themes \cite{watanabe2017lyrisys}, or visual images as descriptors \cite{watanabe2023text}, could enhance both the quality and diversity of the generated lyrics.
However, we only employ this method for fine-tuning to make BERT models learn the structure and moral expressions in lyrics. The creation of truly creative lyrics for artistic purposes requires greater sophistication and rigorous human review \cite{watanabe2023text}. 
Further, we analysed the overall moral expressions in the song lyrics without differentiating between structural elements such as verses, bridges, and choruses. 
Lastly, we focus on inferring moral values in English lyrics, which limits our ability to understand moral expressions in music lyrics from non-Western cultures.

Understanding how lyrics can convey moral values is important for the MIR field, as it can enhance how we experience and interact with music, including improving music tagging and recommendation systems \cite{laplante2014improving}.  
Addressing challenges in automatic detection of moral values in lyrics can further push the boundaries of current technologies in natural language processing and machine learning applied to music and other creative tasks.
Further, as lyrics often reflect societal values and cultural norms, tools for extracting morality rapidly from lyrical text enable researchers to gain insights into the prevailing moral attitudes of different times or cultures. This can be useful in sociological studies, helping scholars understand how music influences and is influenced by societal norms and changes.

\section{Ethics Statement}
In this study, we employed large language models (LLMs) to generate synthetic lyrics. Given the vast amount of data on which these models are trained, there is a potential for bias transfer from the training datasets. Additionally, these models may inadvertently contain copyrighted literary works within their training data, necessitating meticulous steps to prevent plagiarism, particularly if the generated lyrics are utilised beyond fine-tuning for artistic and creative outputs \cite{barnett2023ethical,morreale2023data}.

We engaged two human annotators to label 200 songs with moral values based on the Moral Foundations Theory (MFT). These annotators signed a consent document that detailed the project's objectives, their roles, and the nature of their tasks. They were informed of their right to withdraw from the study at any time without consequences. To protect their privacy, all data from the annotators were anonymised.

While powerfull language models like BERT and GPT-4 offer significant potential to enhance communication and support social campaigns, they also pose risks if used for manipulative purposes. Our research is committed to advancing the understanding of moral expressions in music and fostering the responsible development and use of AI in creative contexts. 


%
%
%
%
%


\end{document}